\documentclass[%
 reprint,
 amsmath,amssymb,
 aps,
 prl,
 superscriptaddress,
]{revtex4-1}
\usepackage[LGR,T1]{fontenc}
\usepackage[latin9]{inputenc}
\usepackage{textcomp}
\usepackage{graphicx}
\usepackage[colorlinks]{hyperref}
\usepackage{lineno}
\usepackage{esint}
\usepackage{color}
\usepackage{colortbl}
\usepackage[english]{babel}
\usepackage{lineno}

\ProvideTextCommand{\~}{LGR}[1]{\char126#1}

\makeatother

\date{\today}

\makeatletter
\definecolor{Gray}{gray}{0.9}
\begin{document}

\title{Dynamics of a Persistent Insulator-to-Metal Transition in Strained
Manganite Films}

\selectlanguage{english}%

\author{Samuel W. Teitelbaum$^{\S}$}
\email{samuelt@alum.mit.edu}
\affiliation{Department of Chemistry, Massachusetts Institute of Technology, Cambridge MA 02139, USA}
\thanks{These authors contributed equally to this work}

\author{B. K. Ofori-Okai$^{\S}$}
\email{boforiokai@alum.mit.edu}
\affiliation{Department of Chemistry, Massachusetts Institute of Technology, Cambridge MA 02139, USA}
\thanks{These authors contributed equally to this work}

\author{Yu-Hsiang Cheng}
\affiliation{Department of Electrical Engineering and Computer Science, Massachusetts Institute of Technology, Cambridge MA 02139, USA}

\author{Jingdi Zhang}
\affiliation{Department of Physics, University of California San Diego, La Jolla, CA, USA}

\author{Feng Jin}
\affiliation{Hefei National Laboratory for Physical Sciences at Microscale, University of Science and Technology of China, Hefei, Anhui 230026, China}
\affiliation{High Magnetic Field Laboratory of the Chinese Academy of Sciences, University of Science and Technology of China, Hefei, Anhui 230026, China}

\author{Wenbin Wu}
\affiliation{Hefei National Laboratory for Physical Sciences at Microscale, University of Science and Technology of China, Hefei, Anhui 230026, China}
\affiliation{High Magnetic Field Laboratory of the Chinese Academy of Sciences, University of Science and Technology of China, Hefei, Anhui 230026, China}
\affiliation{Institutes of Physical Science and Information Technology, Anhui University, Hefei 230601, China}

\author{Richard D. Averitt}
\affiliation{Department of Physics, University of California San Diego, La Jolla, CA, USA}

\author{Keith A. Nelson}
\email{kanelson@mit.edu}
\affiliation{Department of Chemistry, Massachusetts Institute of Technology, Cambridge MA 02139, USA}

\begin{abstract}

Transition metal oxides possess complex free energy surfaces with competing degrees of freedom. Photoexcitation allows shaping of such rich energy landscapes. In epitaxially strained $\mathrm{La_{0.67}Ca_{0.33}MnO_3}$, optical excitation with a sub-100 fs pulse above $2\ \mathrm{mJ/cm^2}$  leads to a persistent metallic phase below 100 K. Using single-shot optical and terahertz spectroscopy, we show that this phase transition is a multi-step process. We conclude that the phase transition is driven by partial charge order melting, followed by growth of the persistent metallic phase on longer timescales.  A time-dependent Ginzburg-Landau model can describe the fast dynamics of the reflectivity, followed by longer timescale in-growth of the metallic phase.

\end{abstract}

\maketitle

\begin{figure}
\includegraphics[width=\columnwidth]{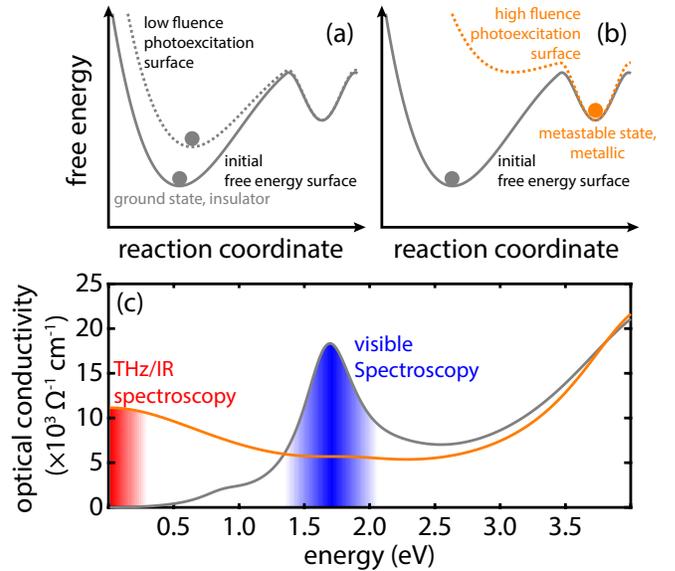}
\caption{(a, b) Schematic illustrations of the minima of the \emph{free energy} surfaces of LCMO. Irradiation changes the initial surface (solid) into the photoexcited surface (dashed) below (a) and above (b) the switching threshold. (c) Optical conductivity of the insulating (grey) and metallic (orange) phases at 100 K extracted from ellipsometry. The blue shaded region of the spectrum is accessible by optical spectroscopy while the red shaded region is accessible by THz/IR spectroscopy.} \label{fig:Switching-schematic}
\end{figure}

Optical control of complex materials provides an opportunity for understanding how strongly interacting degrees of freedom couple to each other to form far-from-equilibrium states with distinct order. By using pulsed laser excitation to control the material's free energy surface, one can reach phases with properties that are not thermally accessible \cite{Tokura2006d,Nasu2001}. Transition metal oxides with competing order are particularly attractive targets for optical control.  These materials have complex energy landscapes with multiple local minima on the free energy surface. Near the phase boundary between two ordered phases, evidence of phase coexistence \citep{Tokura2006,Dagotto2001,Lai2010} offers promise for using light to switch between two ordered phases (Fig. \ref{fig:Switching-schematic}(a),(b)). If light can reshape the free-energy surface and drive the oxide crystal into new local minimum, a metastable phase that does not exist at equilibrium can form \cite{Stojchevska2014,Stoica2019}. 

The strong interaction of lattice, spin, and orbital degrees of freedom in manganite oxides results in a rich phase diagram \cite{Tokura2006,Dagotto2005}. Tuning the orbital overlap between neighboring \textit{d}-electrons in the crystal (the electronic bandwidth) modifies the energetic balance between insulating and metallic phases on the phase diagram and the effective intersite magnetic exchange interaction.  When these two phases are nearly degenerate, the small free energy difference results in photosensitive materials that switch back and forth upon photoexcitation. One approach to controlling the orbital overlap is with chemical pressure -- changing the size of the cations in the perovskite structure to tune the Mn-O-Mn bond angle, which in turn tunes the electronic bandwidth, \emph{e.g.} in  $\mathrm{Pr_{0.55}(Ca_{1-y}Sr_{y})_{0.45}MnO_3}$, where $0.2<y<0.24$. In this compound, photoexcitation with femtosecond pulses results in long-lived photodarkening at low fluence, and permanent photodarkening at high fluence which eventually results in a macroscopic, ferromagnetic phase \cite{Matsubara2008}.

An alternative approach to tuning the Mn--O--Mn bond angle is epitaxial strain. While bulk $\mathrm{La_{0.67}Ca_{0.33}MnO_{3}}$ (LCMO) is metallic and ferromagnetic at low temperature, thin LCMO on $\mathrm{NdGaO_3}$ (001) (NGO) is insulating and antiferromagnetic \cite{Huang2012, Wang2013b}. Excitation of the LCMO film with femtosecond near infrared pulses at high fluence results in a metastable ``hidden'' ferromagnetic phase, which persists until the film is heated to above 140 K \cite{Zhang2015}. These observations imply that there is a region of the manganite phase diagram where the delicate balance of competing orders conspires to generate two nearly degenerate phases, and that it is possible to switch between them using light. 

Questions remain about what steps after the initial excitation event lead to a persistent metallic phase, and the role of lattice vibrations, strain, electronic excitation, and magnetic domain formation in the creation of the metallic phase.  In the framework of Figs. \ref{fig:Switching-schematic}(a) and \ref{fig:Switching-schematic}(b), what are the components of the ``reaction coordinate'' that stabilizes the metallic phase? Considerable insight could be inferred from knowledge of the phase transition dynamics, but obtaining that information is challenging because upon photoswitching, the material does not return to the insulating phase after each laser shot. This means traditional, stroboscopic time-resolved pump-probe spectroscopy where the sample returns to its initial state after every laser shot is not a viable method for investigating the switching process. 

We used single-shot time-resolved spectroscopy in the near-infrared and terahertz (THz) spectral ranges to observe depletion of the charge-ordered, insulating phase and appearance of the metallic phase in strain-engineered LCMO. Our measurements were made on a 36 nm thick LCMO film grown on NGO $(001)$. The steady-state change in DC conductivity and optical properties can be described by spectral weight transfer from an intersite $d-d$ transition band, centered at 1.7 eV, to a Drude peak in the far-infrared (Fig. \ref{fig:Switching-schematic}(c)). The key results of this work are the time-dependent conductivity and optical reflectivity measured over several hundred ps after an above-threshold excitation pulse. After photoexcitation above the switching threshold, we observe a rapid depletion of the charge-transfer band, while the rise of Drude peak is slow. This is attributed to a multi-step process in which the phase transition is driven by sudden excitation of a high density of intersite $d-d$ charge transfer excitations. These relax to form metallic regions that give rise to a DC conductivity over the nanosecond-to-microsecond timescale as the metallic phase forms.

All photoswitching measurements were performed on samples that were cooled to a temperature of 80 K. Following photoexcitation by just one laser pulse above the switching threshold, the samples were reset by heating above 130 K and cooling to 80 K. Measurements of the optical reflectivity and THz transmission confirmed that the sample returned to the pristine state after temperature cycling. Monitoring the persistent change in visible reflectivity and THz conductivity after photoexcitation reveals the fluence required for switching and the fraction of material switched. Below the threshold fluence of 2 mJ/cm$^2$, no permanent change was observed. Above this threshold, the charge-ordered phase is depleted, with more depletion at higher fluence. At a fixed fluence, irradiation with subsequent pump pulses did not increase the amount of switched material. This was observed by spectral weight transfer from the intersite $d$-$d$ peak at 1.7 eV to the far-infrared spectral region. The appearance of the metallic phase can be inferred from the DC conductivity, as extracted from THz time-domain spectroscopy.

\begin{figure}
\includegraphics{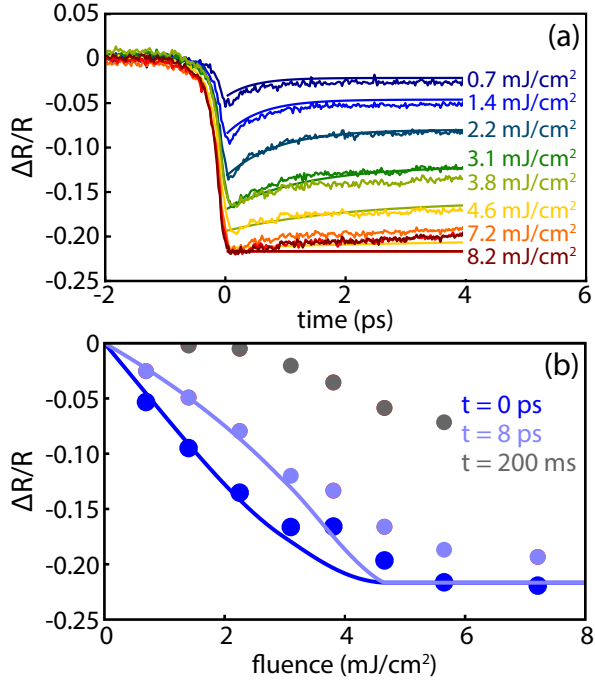}
\caption{Single-shot optical spectroscopy of a 36 nm film of LCMO. (a) Single-shot transient reflectivity as a function of incident fluence. Thick lines are experiment, and thin lines are derived from a time-dependent Ginzburg-Landau model described in the text.  (b) Transient reflectivity at $t = 0$ as a function of excitation fluence.  Solid lines are derived from the time-dependent Ginzburg-Landau model.  Reflectivity values at 8 ps and 200 ms are also shown, revealing rapid partial recovery to a level that persists beyond 300 ps and ultimate recovery to a level that remains constant after 1 ms.}
\label{fig:Single-shot-results}
\end{figure}

To monitor the temporal evolution of the charge order via the change in the optical reflectivity at 1.5 eV photon energy, we used dual-echelon single-shot spectroscopy \citep{Shin2014} to record the real-time  reflectivity change in LCMO at a probe energy of 1.5 eV over a 9 ps window. The relative reflectivity changes, $\Delta R/R$, after a single shot for a series of excitation fluences are shown in Fig. \ref{fig:Single-shot-results}(a).

The initial reflectivity decrease in the near-infrared is caused by a decrease in the amplitude of the intersite $d$-$d$ transition amplitude due to partial (or nearly complete) melting of the charge ordered phase \citep{Beaud2014,Matsubara2008}. We used time-dependent Ginzburg-Landau theory \citep{Beaud2014,Esposito2018} to model the time-dependent reflectivity, similar to other treatments of charge density wave melting \citep{Huber2014,Yusupov2010,Rettig2019}. In the case of LCMO, the time and depth-dependent energy density, $n(t,z)$, following photoexcitation can be described as

\begin{equation}
n(t,z) = (n_e(z) - \alpha n_c) e^{-\Gamma t} + \alpha n_c, \label{eq:TD-GL-theory}
\end{equation}

\noindent where $n_e(z)$ is proportional to the depth-dependent initial energy density.  $\alpha = 1-(1-n_{e}(z)/n_{c})^{2 \gamma}$. The time-dependent fraction of material in the charge-ordered state is $X(t,z) = 1-n(t,z)/n_c$ when $n < n_c$.  The adjustable parameters of this model are the overall short-time decay rate $\Gamma$, the critical exponent $\gamma$, and the critical density $n_c$.  From our fits, we find $n_c = 0.09$ absorbed photons per Mn atom, corresponding to an incident fluence of 3 mJ/cm$^2$, $\Gamma = 2.7(n_c-n_e)/n_c$ ps$^{-1}$, and $\gamma = 0.25$.  Finally, we used an effective medium dielectric function model along with a multilayer reflectivity model to calculate the change in reflectivity of the thin film.  The time-dependent reflectivity calculated from this model is shown in Fig. \ref{fig:Single-shot-results}(a) as thin solid lines. Agreement between the saturation reflectivity at high fluence in Fig. \ref{fig:Single-shot-results}(a) and the model was found when we allowed only the first 30 nm of the 36 nm film to respond to light.  

In order to quantify the fraction of material driven into the metallic phase, we used a model based on spectral weight transfer and effective medium theory. We treat the switched LCMO film as a combination of insulating and metallic LCMO at 80 K. The material has an effective dielectric constant determined by a weighted average of the dielectric functions of the insulating and metallic phases, with volume fractions $1-X$ and $X$ respectively. The dielectric functions for each phase can be described as a sum of four harmonic oscillators,

\begin{equation}
\varepsilon_{i,m}(\omega)=\varepsilon_{\infty}+\sum_{n=1}^4\frac{\omega_{p,n}^{2}}{\omega_{0,n}^{2}-\omega^{2}-i\gamma_{n}\omega}, 
\label{eq:Optical-Conductivity-Model}
\end{equation}

\noindent where $\omega_{p,n}$, $\omega_{0,n}$, and $\gamma_{n}$, are the plasma frequency, resonant frequency, and damping rate for the $n$-th oscillator, respectively. Parameter tables for this model are given in the supplemental information \footnote{See supplemental information for tables of the dielectric funtion model and additional details of the THz conductivity measurements}.  

Though our approach is similar to the treatment used in ref. \citep{Beaud2014}, our data requires a fluence-dependent relaxation rate $\Gamma$ that diverges as the charge order melts.  We note that although the model accurately reproduces the data at short times, there is a discrepancy at longer time delays at high fluence, as shown in Fig. \ref{fig:Single-shot-results}(b). We observed that the metastable metallic state possesses slightly different optical properties compared to the bulk metallic state. As such, we expect discrepancies when using the dielectric function bulk metallic state parameters to describe the mixed insulator-metastable-metallic state. Even so, the agreement is sufficiently good to suggest that the picture of charge order melting formulated in previous studies on insulating manganites describes the temporal evolution of optical reflectivity that we observe at early times. However this treatment does not account for effects such as the renormalization of the intersite exchange interactions expected during an antiferromagnetic-to-ferromagnetic transition.

\begin{figure}
\includegraphics[width=\columnwidth]{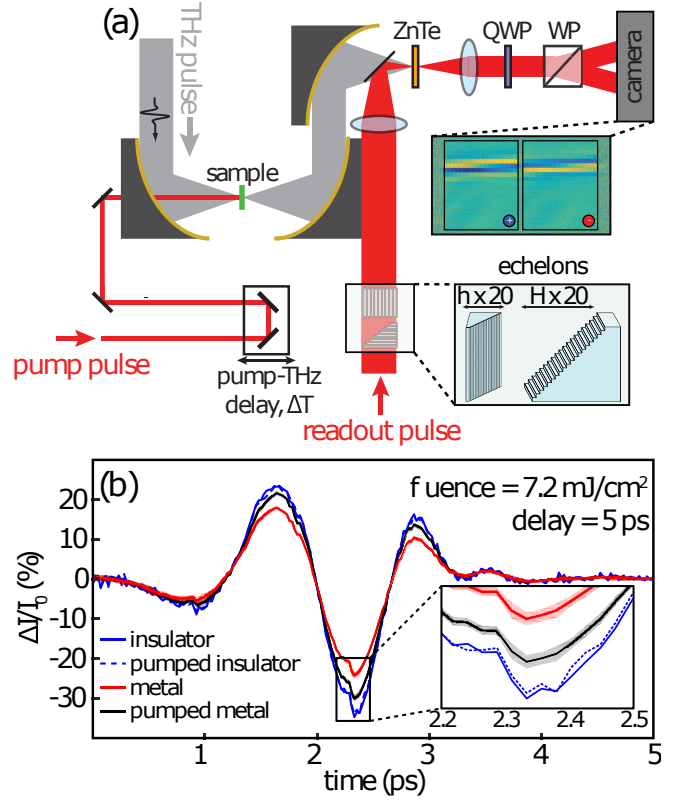}
\caption{(a) Schematic illustration of the optical pump-THz probe setup. The pump pulse was time-delayed using a mechanical stage and overlapped with the THz pulse through a hole in an off-axis parabolic reflector. The readout pulse was split into a 20$\times$20 grid of time-delayed pulses using two transmissive echelons. The pulses were overlapped with the transmitted THz field in a ZnTe crystal. The THz-induced birefringence was analyzed using a quarter wave plate (QWP) and Wollaston prism (WP). The inset shows a computed ratio image (THz pulse present/THz pulse absent) of orthogonal, (+) and (-), polarization states of the readout beam. (b) Single-shot THz time-domain waveforms of the static and pumped insulating and metallic films. The inset shows the difference between the pumped metal and pumped insulator.} \label{fig:Thz-single-shot}
\end{figure}

The change in quasi-DC conductivity (\emph{i.e.} itinerant carrier density) upon entering the metallic phase was measured by single-shot THz spectroscopy \citep{Teo2015, Ofori-Okai2018, Katayama2019}. This records the THz time-domain waveform over a 9.28 ps window in a single laser pulse. As illustrated in Fig. \ref{fig:Thz-single-shot}(a), the readout pulse is passed through two 20-step echelon structures with step heights of $h = 15$ $\mu$m and $H = 300$ $\mu$m. The echelons impose time delays of 23.1 fs and 462 fs per step, respectively, to produce a 20$\times$20 grid of pulse. The pulses are overlapped with the transmitted THz signal in the electro-optic ZnTe crystal. Via the Pockels effect, the polarization rotation of each pulse is proportional to the THz field at a unique time point. The readout pulses are passed through a quarter-wave plate (QWP) and split by a Wollaston prism (WP) for balanced detection of orthogonal polarizations at the CMOS camera.  Example THz time-domain waveforms are presented in Fig. \ref{fig:Thz-single-shot}(b) showing the THz pulse transmitted through the pristine insulator (solid blue), the photo-switched metal (solid red), the photoexcited insulator (dashed blue), and the photoexcited metal (black). The gray area around the black line shows the standard deviation over 50 shots, giving an estimate of the uncertainty in the measurement. For the measurements shown here of the photoexcited insulator and metal, the excitation pulse arrived 5 ps before the THz probe pulse. From the measured THz transmission, the conductivity was extracted as described in \citep{Tinkham1956} and in the supplemental material. For longer time delays, optical pump-THz probe measurements on the switched film showed results consistent with earlier measurements \cite{Averitt2001}.

\begin{figure}
    \centering
    \includegraphics[width=\columnwidth]{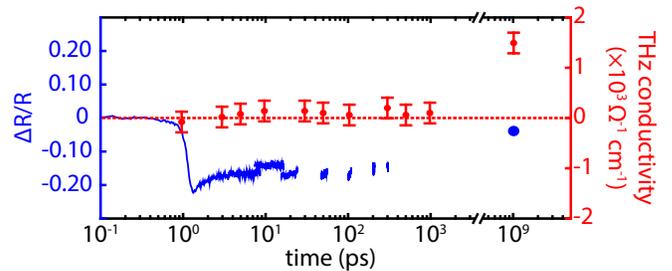}
    \caption{Single-shot transient reflectivity at 800 nm and THz conductivity as a function of time after above-threshold excitation with fluence of $>$7 mJ/cm$^{2}$, along with the steady-state values reached 5 ms after photoexcitation. The dashed horizontal line shows zero conductivity.}
    \label{fig:Single-shot-results_followup}
\end{figure}

A comparison between the time-dependent THz conductivity (shown in red) and optical reflectivity (shown in blue) is presented in Fig. \ref{fig:Single-shot-results_followup}. Longer delays were obtained by using a mechanical stage to vary the arrival time of the excitation pulse relative to the echelon time window, for optical reflectivity or THz conductivity measurements. After photoexcitation above the switching threshold, we observe a rapid ($<$30 fs) depletion of the inter-site hopping band, while the rise of Drude peak is slow ($>$1 ns). For an excitation fluence of 7.2 mJ/cm$^{2}$, the observed change in THz conductivity at short times (<1 ns) is $\sim 100\ {\Omega^{-1} \mathrm{cm}}^{-1}$, far lower than the conductivity of the persistent metallic phase of 3500 $\Omega^{-1} \mathrm{cm}^{-1}$\citep{Zhang2015}, and distinct from the conductivity of the metallic phase following photoexcitation \cite{Averitt2001}. This is further supported by comparing the dashed blue line in Fig. \ref{fig:Thz-single-shot} with the black line and the gray area around it. By 1 ms after photoexcitation, the conductivity has risen to 1,500 $\Omega^{-1} \mathrm{cm}^{-1}$, and no further change is observed. Figure \ref{fig:Single-shot-results_followup} also shows the optical pump-probe trace over longer timescales. The change in static reflectivity confirmed that the switching is consistent for each photoexcited spot. For all optical measurements, the sample reflectivity drops by 20\%, and remains low for the longest times investigated (300 ps) before the next laser shot approximately 200 ms later, at which point it recovers to an 8\% drop and does not change with further photoexcitation.  The partial recovery of the optical reflectivity to its equilibrium value and the intermediate conductivity in the THz regime indicate that only a fraction of the sample has been switched into the metallic phase.

Although the fast changes in optical reflectivity in the 1.5-3 eV range at early times are described well by the picture of charge-order melting, the corresponding changes in the THz conductivity are much slower and not indicative of a state that is a mixture of the insulating and metallic phases. This suggests that the metallic phase is not immediately formed upon suppression of charge order by photoexcitation. If it were, the conductivity would increase on the timescale of the electron-phonon coupling, a few hundred fs \cite{Wall2009}. The single-shot THz spectroscopy measurements show that the onset of macroscopic conductivity takes more than 1 ns.

Instead, the discrepancy in timescale between the depletion of the inter-site hopping band in the visible range and the onset of THz conductivity indicates that there is a long-lived intermediate state that slowly evolves into the metallic phase. As the optical conductivity change requires only inter-site hopping of electrons, which are thought to be localized to approximately the size of one unit cell \cite{Ahn2003,Ahn2013}, charge order melting can account for the observed optical reflectivity changes. The strong electron-phonon coupling leads to a change in the localized lattice distortions around specific excitation sites. This highly excited insulating state must then reorganize into a stable metallic phase containing domains that are larger than the percolation threshold \citep{Walther2007} in order to produce a macroscopically conducting material, on a timescale longer than 1 ns. In addition, the magnetic ordering must shift from antiferromagnetic to ferromagnetic, concomitant with an underlying evolution of the intersite exchange interactions. This is consistent with the measured discrepancy in timescales of evolution of the optical reflectivity and THz conductivity. This picture is also consistent with speculation first proposed in ref. \cite{Matsubara2008}. However, the work in ref. \cite{Matsubara2008} lacked real-time measurements above the persistent switching threshold, whereas our work provides evidence of long timescale (>1 ns) dynamics that are critical to understanding the phase transition.

We conclude that the onset of the hidden phase in LCMO is a multistep process requiring several relaxation steps that produce a stable, conductive, ferromagnetic phase following photo-induced collapse of the insulating charge-ordered phase. In the first step, a significant fraction ($>$1/2) of the charge order is quickly suppressed by photoexcitation. In the second step, the local Jahn-Teller distortions around the Mn sites relax \citep{Wu2009,Li2016,Prasankumar2007a}, melting the charge ordered structural phase and favoring a state with local ferromagnetic coupling. Finally, locally ferromagnetic regions aggregate into larger ferromagnetic domains on nanosecond (or longer) timescales, which are stabilized by long range strain and ferromagnetic order. Once the metallic domain size reaches the percolation threshold in the THz spectral range, the observed conductivity rises as the optical reflectivity partially recovers. This picture is consistent with the large roles local structure, phase segregation, and local strain play in the photoinduced dynamics of manganites, and accounts for the difference in timescales between optical and THz single-shot observations.  

\begin{acknowledgments}
S.W.T, B.O.O. Y.H.C. and K.A.N acknowledge  support from the Office of Naval Research Grants No. N00014-12-1-0530 and No. N00014-16-1-2090 the National Science Foundation Grant No. CHE-1111557, and the Samsung GRO program.  J. Z. and R.D.A. acknowledge support from DOE - Basic
Energy Sciences under Grant DE-SC0012375.  F.J. and W.W. acknowledge supports from the National Basic Reasearch program of China Grant No. 2016YFA0401003 and Hefei Scienence Center CAS.
\end{acknowledgments}

\bibliographystyle{apsrev4-1}
\bibliography{Citations_for_LCMO_Paper.bib}

\end{document}


\title{Supplemental Information for: Dynamics of a Persistent Insulator-to-Metal Transition in Strained
Manganite Films}

\author{Samuel W. Teitelbaum$^{\dagger}$}
\email{samuelt@alum.mit.edu}
\affiliation{Department of Chemistry, Massachusetts Institute of Technology, Cambridge MA 02139, USA}

\author{Benjamin K. Ofori-Okai$^{\dagger}$}
\email{benofori@mit.edu}
\affiliation{Department of Chemistry, Massachusetts Institute of Technology, Cambridge MA 02139, USA}

\author{Yu-Hsiang Cheng}
\affiliation{Department of Electrical Engineering and Computer Science, Massachusetts Institute of Technology, Cambridge MA 02139, USA}

\author{Jingdi Zhang}
\affiliation{Department of Physics, University of California San Diego, La Jolla, CA, USA}

\author{Feng Jin}
\affiliation{Hefei National Laboratory for Physical Sciences at Microscale, University of Science and Technology of China, Hefei, Anhui 230026, China}
\affiliation{High Magnetic Field Laboratory of the Chinese Academy of Sciences, University of Science and Technology of China, Hefei, Anhui 230026, China}

\author{Wenbin Wu}
\affiliation{Hefei National Laboratory for Physical Sciences at Microscale, University of Science and Technology of China, Hefei, Anhui 230026, China}
\affiliation{High Magnetic Field Laboratory of the Chinese Academy of Sciences, University of Science and Technology of China, Hefei, Anhui 230026, China}
\affiliation{Institutes of Physical Science and Information Technology, Anhui University, Hefei 230601, China}

\author{Richard D. Averitt}
\affiliation{Department of Physics, University of California San Diego, La Jolla, CA, USA}

\author{Keith A. Nelson}
\email{kanelson@mit.edu}
\affiliation{Department of Chemistry, Massachusetts Institute of Technology, Cambridge MA 02139, USA}

\maketitle

\section{Methods}

Static conductivity measurements were performed using a conventional THz time-domain spectroscopy (THz-TDS) setup. The THz pulses were produced by optical rectification in a lithium niobate crystal using 1 mJ pulses, and were collected and focused into the samples using a pair of off-axis parabolic reflectors. The THz focal spot was $\sim1$ mm diameter at the sample. The waveform was measured using conventional electro-optic (EO) sampling in a 1 mm ZnTe crystal. The excitation pulse was directed onto the sample at normal incidence through a 3 mm hole in the off-axis parabolic reflector. This ensured that that the optical pump pulse was larger than the THz probe pulse. The pulse energy was attenuated using a half waveplate and polarizer, and was blocked downstream of the sample using a Teflon beam block. As mentioned in the main text, the sample was reset by temperature cycling after each measurement of the photoswitched material.

From the time-domain spectroscopy measurements, the conductivity was determined from the experimentally measured THz transmission spectrum, $\tilde{t}(\omega)$ using the Tinkham equation,
\begin{equation}
\tilde{\sigma}(\omega) = \frac{n_{s}+1}{Z_{0}d}\left(\frac{1}{\tilde{t}(\omega)}-1\right),
\end{equation}

\noindent where $n_{s}=4.7$ was the refractive index of the NGO substrate, also measured by THz-TDS, $Z_{0}=377$ $\Omega$ is the impedance of free space, and $d$ is the film thickness. 

The single-shot THz measurements were performed using dual-echelons for EO sampling \citep{Teo2015}. The echelons generated the time-delayed optical pulses needed to readout the THz waveform. They were focused into the ZnTe crystal and then subsequently imaged onto a scientific CMOS (ANDOR Zyla 5.5, 2560 $\times$ 2048 pixels, 16 bit, 30 Hz maximum full-chip acquisition rate) with a $1/2\times$ magnification. Polarization gated balancing was accomplished using a quarter waveplate and small-angle splitting Wollaston prism. As with the single-shot optical measurements, the number of pulses was limited using a mechanical shutter and external Pockels cell.

\section{Transient Reflectively Model}

Tables \ref{tab:Paramater-table-for-CO-band} and \ref{tab:Paramater-table-for-CO-band-ES} show the parameters for the effective medium model used in this work. Highlighted rows are sensitive to photoexcitation.

\begin{table}[ht]
\begin{tabular}{|c|c|c|c|}
\hline 
$n$ & $\omega_{0}$ (eV) & $\omega_{p}$ (eV) & $\gamma$ (eV)\tabularnewline
\hline 
\hline 
1 & 0.888 & 0.6552 &  0.4477\tabularnewline
\hline 
\rowcolor{Gray}
2 & 1.691 & 2.749 & 0.496\tabularnewline
\hline 
3 & 4.226  & 5.258 & 1.872\tabularnewline
\hline 
4 & 8.564 & 15.351 & 5.219\tabularnewline
\hline 
\end{tabular}

\caption{Parameter table for the dielectric function of LCMO in the charge ordered (insulating) phase.
\label{tab:Paramater-table-for-CO-band}}

\end{table}

\begin{table}[ht]
\begin{tabular}{|c|c|c|c|}
\hline 
$n$ & $\omega_{0}$ (eV) & $\omega_{p}$ (eV) & $\gamma$ (eV)\tabularnewline
\hline 
\hline 
1 & 0 & 0.3555 & 1.213\tabularnewline
\hline 
\rowcolor{Gray}
2 & 1.899 & 1.00 & 0.987\tabularnewline
\hline 
3 & 4.142 & 4.616 & 1.41\tabularnewline
\hline 
4 & 8.564 & 15.35 & 5.21\tabularnewline
\hline 
\end{tabular}

\caption{Parameter table for the dielectric function of LCMO in the ferromagnetic (metallic) phase.
of LCMO.\label{tab:Paramater-table-for-CO-band-ES}}
\end{table}

\begin{figure}
    \centering
    \includegraphics{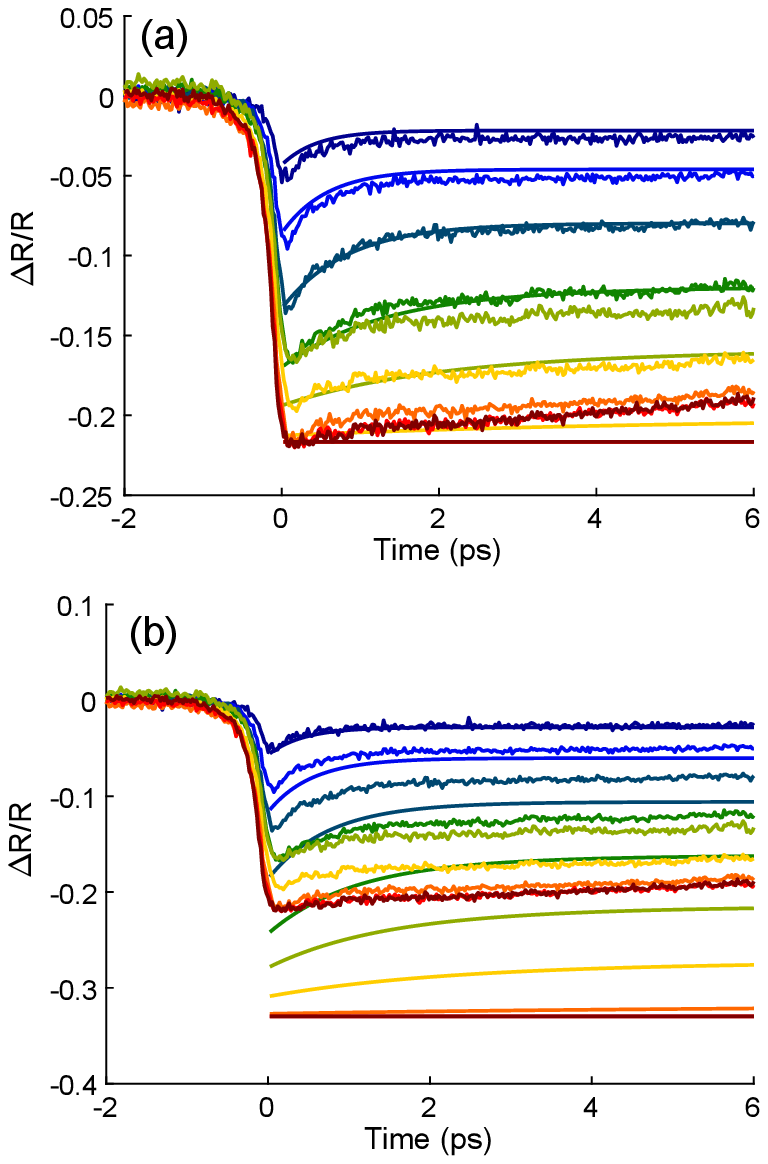}
    \caption{(a) Ginzburg-Landau Modeling using effective medium theory, with adjustable response depth to tune the saturation reflectivity.  (b) Ginzburg-Landau Modeling using effective medium theory with no re-scaling, affecting the whole film.  Depth-dependent switching is shown in fig. 3(a), same as (a) in this figure.}
    \label{fig:SI-modeling-details}
\end{figure}

Following the work of Beaud et al, we construct a time-dependent Ginzburg-Landau model to understand the time-dependent optical reflectivity at early times.   Below we describe this process in more detail.  First, we construct the time-dependent excitation density, as described in the text, and use it to derive a depth-dependent order parameter, and then to the fraction of the material switched.  Starting from eq. 1 in the text:

\begin{equation}
n_e(z)= \frac{F(1-R)}{\delta E_{ph} }e^{-z/\delta}, \label{eq:fluence_conversion-SI}
\end{equation}

\begin{equation}
n(t,z) = (n_e(z) - \alpha n_c) e^{-\Gamma t} + \alpha n_c, \label{eq:TD-GL-theory-SI}
\end{equation}

The dielectric function as a function of $z$ can then be calculated as

\begin{equation}
    \varepsilon(z,t) = \varepsilon_{I}(1-X(z,t)) + \varepsilon_{M}X(z,t)
\end{equation}

Where $\epsilon_I$ and $\epsilon_M$ are the insulating and metallic dielectric functions, respectively. We then use a multilayer propagation matrix reflectivity model with 20 layers plus the substrate dielectric function to calculate the time-dependent reflectivity.  The probe light is incident at a 45 degree angle, $\lambda = 800$ nm, p-polarized.  Fig. \ref{fig:SI-modeling-details} shows the results of the model without depth dependent X (bottom panel), and without depth-dependent X, but an arbitrary rescaling of the overall reflectivity to match the saturation value observed experimentally (top panel).  This model does not assume anything about the size or shape of the mixture, nor whether the new phase is distributed over depth or by lateral domains.  While more sophisticated effective-medium models like the Maxwell-Garnett model can also be used to describe this mixture, there is no significant improvement in agreement between experiment and our simulations when using a more complex model for the dielectric function of the photoexcited state.

\section{Optical-pump THz Probe of switched film}

\begin{figure}[h!]
    \centering
    \includegraphics{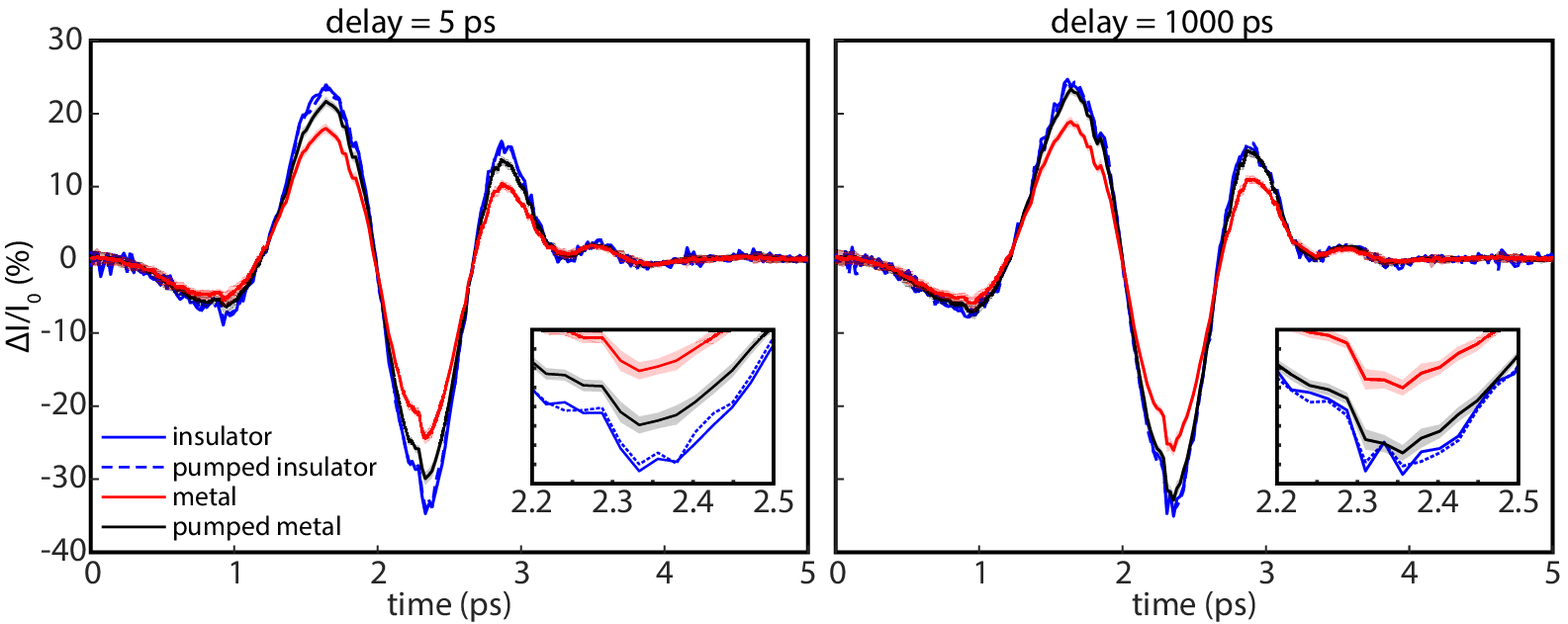}
    \caption{Measured THz waveforms transmitted through the LCMO film in the insulating (solid blue), the photoexcited insulating (dashed blue), switched metallic (solid red), and photoexcited metallic (solid black) lines. In (a), the pump pulse arrives 5 ps before the THz pulse, identical to figure 3 in the main text, and in (b) the pump pulse arrives 1000 ps before the THz pulse.}
    \label{fig:SI-Time_dependence_SS_THz}
\end{figure}

Figure \ref{fig:SI-Time_dependence_SS_THz} shows single-shot THz waveforms collected on the 36 nm LMCO film under different conditions: the pristine insulator (solid blue), the photoexcited insulator (dashed blue), switched metal (solid red), and photoexcited metal (solid black). The time-delay (a) and (b) are 5 ps and 1000 ps, respectively. Figure \ref{fig:SI-Time_dependence_SS_THz}(a) is also shown in the main text. These data illustrate the dynamics of the single-shot THz response following photoexcitation. The data also illustrate the findings of the paper -- the difference in conductivity between the pristine insulator and the photoexcited insulator remains small even at long time delays. In comparison, there is a noticable change in the conductivity of the metallic LCMO upon photoexcitation.

\begin{figure}
    \centering
    \includegraphics{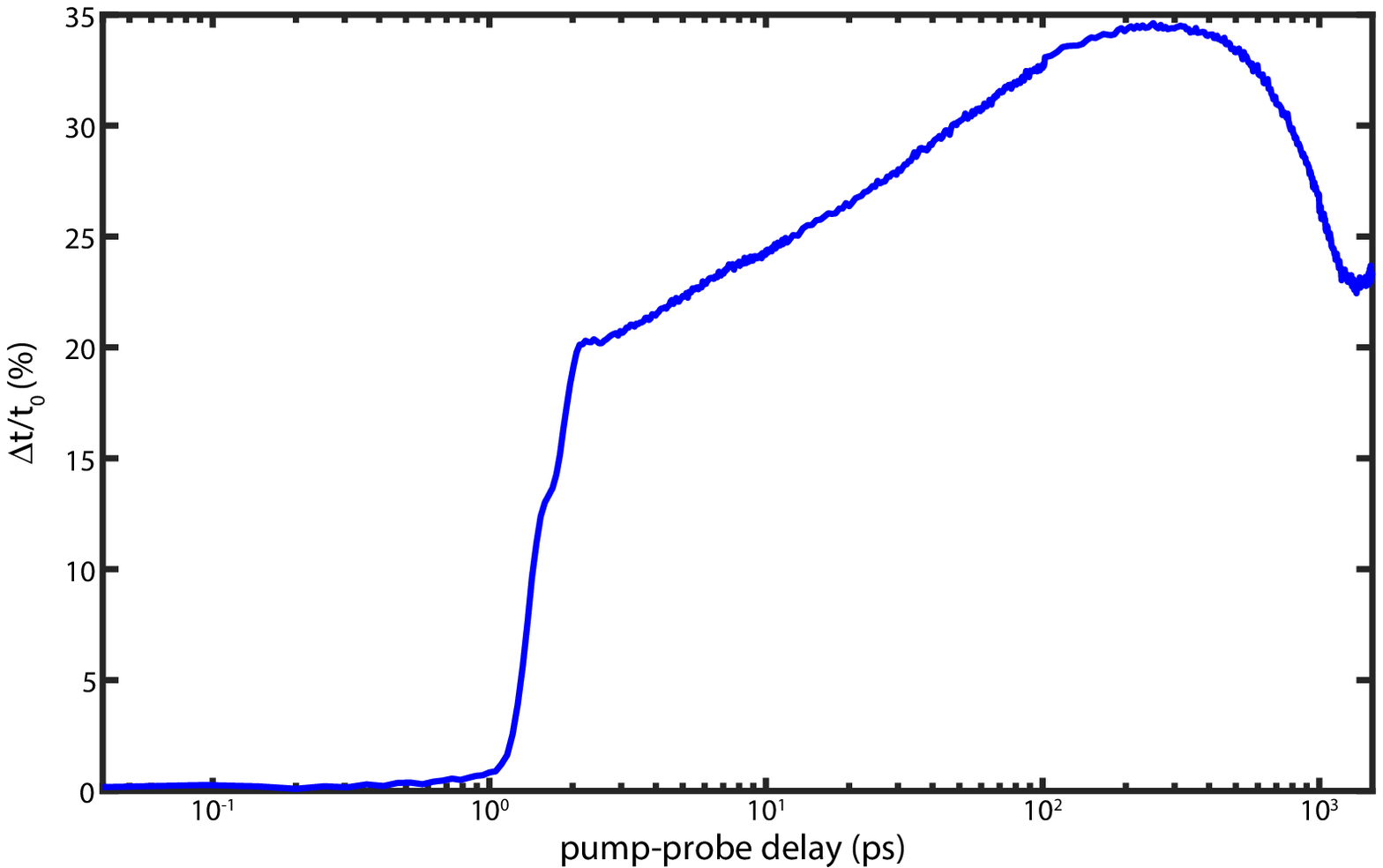}
    \caption{Time-resolved THz measuremens on the 36 nm LCMO film. The data show qualitatively similar behavior to bulk, metallic LCMO on the sub-100 ps timescale.}
    \label{fig:SI_TRTS_LCMO}
\end{figure}

We also confirmed that the behavior of the LCMO film after photoswitching is qualitatively similar to that of bulk LCMO by performing conventional time-resolved THz measurements. A sample dataset are shown in figure \ref{fig:SI_TRTS_LCMO}. Here, the relative change in the transmission at the peak of the THz waveform, $\Delta t/t_{0}$ was measured at a variety of time delays following a $\sim$4 mJ/cm$^{2}$ fluence pulse. Time zero is set to 1 ps so that the data can be plotted on a log scale. As in Ref. \cite{Averitt2001}, we observe an initial fast increase in the THz transmission following photoexcitation, followed by a slower rise at longer timescales. These are consistent with the previously observed heating of the electron system, followed by subsequent heating of the lattice subsystem, through electron-phonon coupling, and spin subsystem, by spin-orbit coupling and momentum scattering. The previous measurements only went out to 100s of ps time delay, while here we measure out to 1.5 ns delay. We observe that after the slow rise there is a similarly slow decay on the THz transmission. This is likely due to heat transfer into the substrate, which leads to cooling of the sample and a rise in the conductivity.

\bibliography{Citations_for_LCMO_Paper.bib}